\documentclass{ar2e}
\usepackage{setspace}
\begin{document}
\input epsf.def   

\input psfig.sty

\jname{Annu. Rev. Nucl. Part. Sci.}
\jyear{2011}
\jvol{1}
\ARinfo{1056-8700/97/0610-00}

\title{Parity-violating Electron Scattering and the Electric and Magnetic Strange Form Factors of the Nucleon}

\markboth{D. S. Armstrong and R. D. McKeown)}{Parity-violating Electron Scattering...}

\author{D. S. Armstrong
\affiliation{Department of Physics, College of William and Mary, Williamsburg, VA 23187}
R. D. McKeown \affiliation{Thomas Jefferson National Accelerator Facility, Newport News, VA 23606;
Department of Physics, College of William and Mary, Williamsburg, VA 23187}}

\begin{keywords}
neutral current, lepton scattering, electromagnetic form factors, strange form factors
\end{keywords}

\begin{singlespace}
\begin{abstract}
Measurement of the neutral weak vector form factors of the nucleon
provides unique access to the strange quark content of the
nucleon. These form factors can be studied using parity-violating
electron scattering. A comprehensive program of experiments has been
performed at three accelerator laboratories to determine the role of
strange quarks in the electromagnetic form factors of the
nucleon. This article reviews the remarkable technical progress
associated with this program, describes the various methods used in
the different experiments, and summarizes the physics results along
with recent theoretical calculations.
\end{abstract}

\maketitle

\section{INTRODUCTION}
Since the initial development of QCD in the 1970's, it has been known
that the internal structure of the nucleon is due to the presence of
quarks, gluons, and a ``sea'' of quark-antiquark pairs. Although the
electric charge of the nucleon is due to the valence quarks (as in the
early quark models of Gell-Mann and Zweig), in QCD the gluons are
critical to quark confinement, generating 98\% of the nucleon mass in
the process. The results of polarized deep inelastic scattering
experiments in the 1980's and 1990's indicated that, contrary to
theoretical expectations, the spin of the nucleon does not arise from
the spins of the quarks. As a result, the role of the gluons and the
quark-antiquark pairs in the static properties of the nucleon became a
subject of great interest. Although the gluons are responsible for
dramatic effects like confinement and nucleon mass, the effects of the
quark-antiquark pairs (necessarily generated by the gluons in QCD and
therefore non-zero) are more difficult to ascertain. One can think of
this as the QCD analog of the famous Lamb shift in atomic physics.

The strange quark-antiquark pairs are of particular interest, since
there are no valence strange quarks in the nucleon and any process
sensitive to strange quarks would necessarily be related to the ``sea".
In 1988, Kaplan and Manohar~\cite{kaplan} proposed the study of study of
strange quark-antiquark pairs by measurements of neutral weak current
matrix elements, perhaps in neutrino scattering experiments. In 1989,
McKeown~\cite{bmck89} and Beck~\cite{beck89} proposed that parity-violating electron scattering
offered a very effective method to study these matrix elements,
generating significant interest and many new experimental
proposals. In the subsequent two decades a great deal of experimental
and theoretical effort resulted in a now rather definitive body of
work that, for the first time, substantially constrains the
contribution of strange quark-antiquark pairs to the elastic
electroweak form factors of the nucleon.

\section{STRANGE QUARKS IN THE NUCLEON}

\begin{figure}
\centerline{\psfig{figure=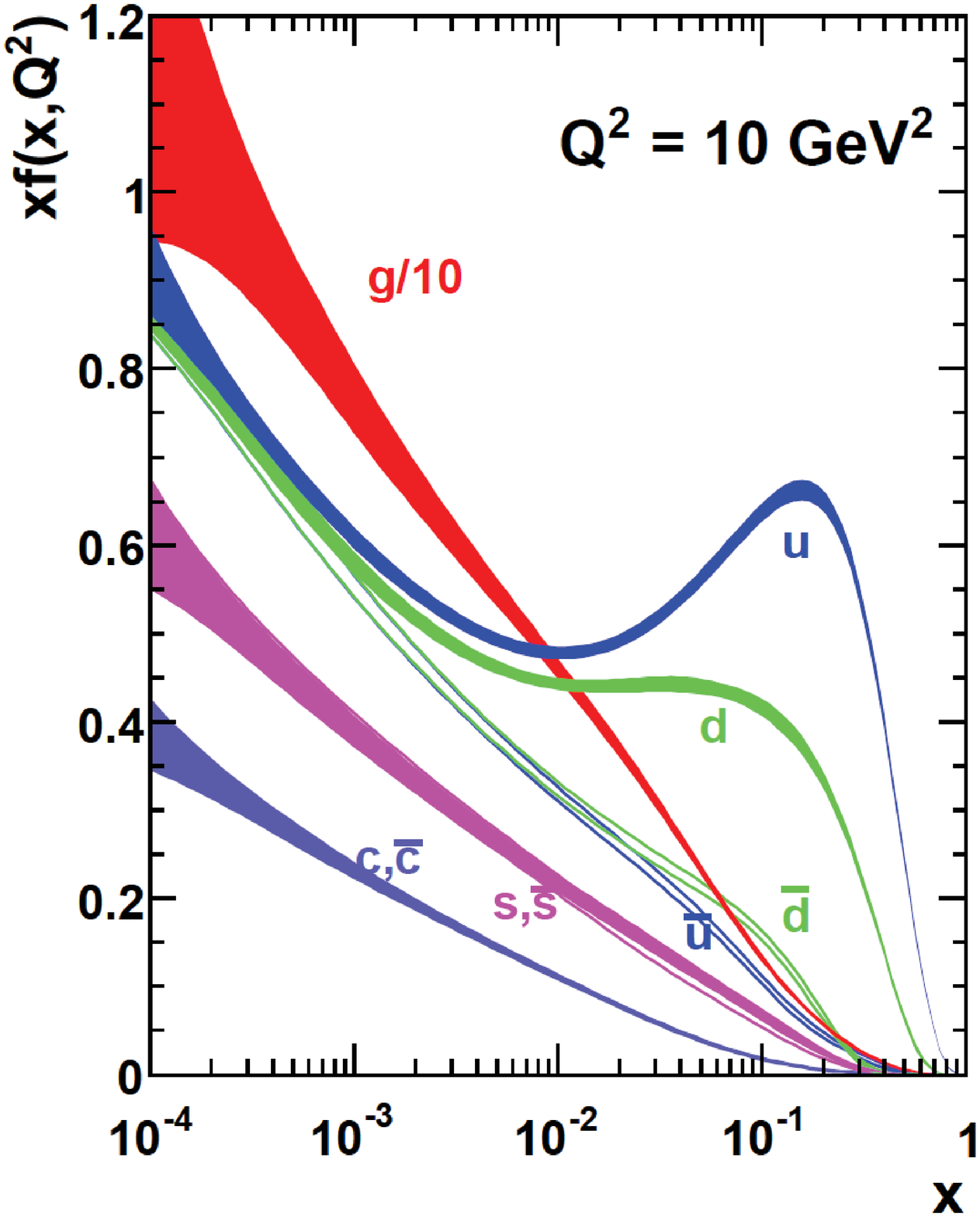,width=4.0in}}
\caption {The results of a global fit \cite{MSTW08} for the quark,
  antiquark, and gluon parton distribution functions as a function of
  the momentum fraction $x$, at $Q^2 = 10$~GeV$^2$. Note that the
  PDF's $f(x)$ are multiplied by $x$ to suppress the large rise at low
  $x$ in the plot, and that the gluon distribution is divided by 10
  for display purposes.}
\label{fig:pdf}
\end{figure}

The most direct evidence for the existence of quarks and antiquarks in
the nucleon is obtained from deep-inelastic lepton scattering. In this
process a high momentum virtual photon interacts with the fundamental
charges in the nucleon (quarks and antiquarks). Through the study of
lepton-proton and lepton-neutron scattering, and using charge symmetry
(with approximate treatment of charge symmetry breaking effects) one
can extract information related to the probability distributions for
up, down, and sea quarks. These distributions, known as parton
distribution functions (PDF), are defined as a function of momentum
fraction in the nucleon, $x$ ($0<x<1$), and are denoted $q(x)$ and
${\bar q} (x)$, where $q= u, d, c, s, t, b$ for the quark flavors. In
addition, the production of like-sign dimuon pairs in deep-inelastic
charged-current neutrino scattering provides information on the
strange quark distribution. Another important ingredient in the flavor
decomposition of the sea quarks is provided by Drell-Yan experimental
data, in which an incident quark (in the projectile nucleon)
annihilates on an antiquark (in the target nucleon) producing a $\mu^+
\mu^-$ pair. Recent Drell-Yan data on the proton and deuteron have
been combined to indicate a startling excess of ${\bar d}(x) / {\bar
  u}(x) $. Finally, the evolution of the quark and antiquark PDF's as
a function of $Q^2$ provides information on the gluon PDF
$g(x)$. Fig.~\ref{fig:pdf} shows the results of the PDF's obtained in
a recent global fit to deep inelastic scattering and Drell-Yan data.

The general features of the PDF's shown in Fig.~\ref{fig:pdf} are
qualitatively well understood. The dominance of $u$ and $d$ at higher
$x>0.1$ indicates the valence nature of these quark flavors (three
valence quarks, each with average $x \sim 1/3$). The dramatic rise of
$g(x)$ at low $x$ is associated with the ``splitting'' of partons at
lower $x$, e.g., $q(x) \rightarrow q(y) + g(x-y)$. The sea quarks
arise from ${\bar q} q$ pair production by gluons via $g(x)
\rightarrow q(y) + {\bar q}(x-y)$. Thus one can infer that the gluons
dominate the dynamics at low $x$ and the presence of the sea quarks is
a secondary feature of the presence of the large number density of
gluons.

However, there is a very interesting and important effect at
intermediate $x$ ($0.01<x<0.1$) where (due to the Drell-Yan data
discussed above) one finds a substantial excess of ${\bar d}(x)$
relative to ${\bar u}(x) $. Clearly the process $g \rightarrow {\bar
  q} q$ would produce ${\bar u} = {\bar d}$ (except for small effects
due to $m_u \ne m_d$). The ${\bar d}(x) / {\bar u}(x) $ excess must
then be attributed to non-perturbative processes. For example, the
fluctuation of a proton into a neutron and $\pi^+$ contains a ${\bar
  d}$ component and there is no symmetric analog process to produce
${\bar u}$. (Note that $p \rightarrow p \pi^0$ produces equal numbers
of ${\bar u}$ and ${\bar d}$). In addition, one would expect that the
${\bar d}$ excess from this process would occur at the value $x \sim
m_\pi / (2 m_N)$, in agreement with the experimental data. Thus this
observed excess of ${\bar d}(x) / {\bar u}(x)$ strongly indicates the
presence of fluctuations into $N \pi$ pairs in the nucleon. Of course,
many models of the nucleon (such as the ``cloudy bag'' models) include
such configurations in an attempt to capture the physics associated
with other observables (such as the anomalous magnetic moments).

Thus one is naturally led to consider the possible role of similar
fluctuations such as $p \rightarrow K^+ \Lambda$. Clearly such
configurations will have lower probability than $N \pi$ fluctuations
due to the higher masses of the $\Lambda$ and $K^+$, but one would
certainly expect there to be finite observable effects resulting from
them.  Such configurations would lead to radially separated
distributions of ${\bar s}$ and $s$ quarks, due to the tendency of the
$K^+$ to occupy larger radial distances from the center of mass of the
$\Lambda K^+$. The spatial separation of the $s$ and ${\bar s}$ would
have several implications:
\begin{enumerate}
  \item ${\bar s} s$ contribution to the nucleon magnetic moment, and other electroweak form factors,
  \item ${\bar s} s$ contribution to the nucleon axial charge, affecting the value of $\Sigma$
associated with the helicity carried by quarks,
  \item ${\bar s} s$ contribution to the mass of the nucleon,
  \item difference between the PDF's $s(x)$ and ${\bar s}(x)$.
\end{enumerate}
The first three of these items are low $Q^2$ or static properties of
the nucleon. Thus they represent a change in the nucleon static
properties analagous to the change in atomic properties (e.g. the Lamb
shift) due to vacuum polarization in QED. The fourth item is an effect
analogous to the ${\bar d}(x) / {\bar u}(x) $ excess observed in the
Drell-Yan process.  In principle, items 1.) and 4.) can be established
in a model-independent fashion, whereas the items 2.) and 3.)
generally require a model-dependent analysis or assumptions about
non-perturbative QCD effects.

\begin{figure}
\centerline{\psfig{figure=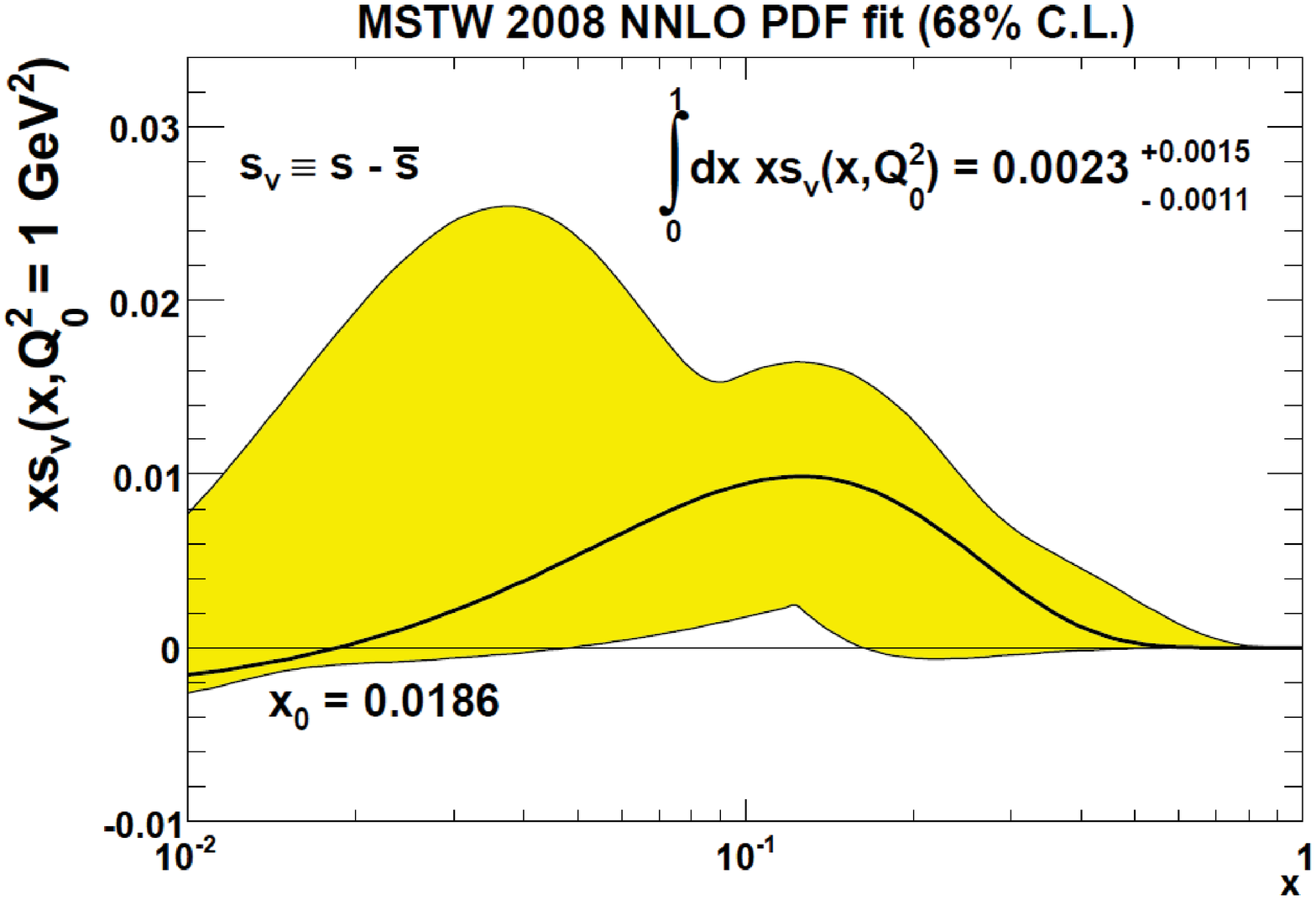,width=4.0in}}
\caption {The results of the recent global fit \cite{MSTW08} for the difference $s(x) - {\bar s} (x)$.}
\label{fig:sminussbar}
\end{figure}

At present, although there have been many hints of effects associated
with items 2-3.) there is no unassailable demonstration that the
evidence can be due to the strange quark effects. For item 4.), there
is only a hint in the latest global PDF fit, MSTW08 \cite{MSTW08}, as
shown in Fig.~\ref{fig:sminussbar}.

For the remainder of this review, we will focus on the theoretical
framework, experimental techniques, and experimental results
associated with the studies of ${\bar s} s$ contribution to the
nucleon magnetic moment, and other electroweak form factors (item 1.)
above).

\section{NEUTRAL WEAK FORM FACTORS}
\label{sec:form}
The elastic scattering of nucleons via the neutral weak current can be described by two vector form factors,
$F_1^Z$ and $F_2^Z$, and an axial vector form factor $G^Z$. These form factors are functions of the invariant
momentum transfer $Q^2$. Like the electromagnetic interaction, the neutral weak interaction with a nucleon
involves coupling to the quarks (and antiquarks) as the gluons have no electroweak interaction. In general,
we can write any of the elastic electroweak form factors in terms of the quark flavors. For example the
electromagnetic form factors are given by
\begin{eqnarray}
  F_1^\gamma &=& \frac{2}{3} F_1^u - \frac{1}{3} F_1^d - \frac{1}{3} F_1^s\\
 F_2^\gamma &=& \frac{2}{3} F_2^u - \frac{1}{3} F_2^d - \frac{1}{3} F_2^s
\end{eqnarray}
where $u$, $d$, and $s$ refer to the up, down and strange quarks. Note that we ignore charm and heavier
quarks as it has been shown that these can be safely neglected.
The neutral weak form factors may be also written in terms of the individual quark flavor components
\begin{eqnarray}
  F_{1,2}^Z &=&(1 - \frac{8}{3} \sin^2 \theta_W) F_{1,2}^u  + (-1+ \frac{4}{3} \sin^2 \theta_W) (F_{1,2}^d + F_{1,2}^s) \\
  G_A^Z &=& -G_A^u + G_A^d +G_A^s
\end{eqnarray}
where $\theta_W$ is the weak mixing angle. The value of $\theta_W$ is, in principle, precisely determined
from other experiments \cite{pdg} although one must consider the renormalization scheme and radiative corrections.

For the vector form factors, one often prefers to use the Sachs form factors
\begin{eqnarray}
G_E^{\gamma, Z} &= & F_1^{\gamma, Z} -\tau F_2^{\gamma, Z} \nonumber \\
G_M^{\gamma, Z} &= & F_1^{\gamma, Z} + F_2^{\gamma, Z}
\end{eqnarray}
with $\tau \equiv Q^2/4 M^2$. In the static limit, the electromagnetic form factors then reduce to the charge
and magnetic moment: $G_E^\gamma (Q^2 = 0) = Q$ and $G_M^\gamma (Q^2 = 0) = \mu$. Another useful quantity is the
charge radius:
\begin{equation}
       \left. \langle r^2\rangle
     = -6 \frac {dG_E(Q^2)}{dQ^2}\right|_{Q^2=0} \ .
\end{equation}

Assuming isospin symmetry, additional relations among the form factors are obtained
using the fact that the transformation of proton to neutron $(p \rightarrow n)$ is equivalent to changing $u$ to $d$
and vice versa $(u \leftrightarrow d)$. For example, the proton and neutron axial form factors are then related to the
quark components (defined for the proton) by
\begin{eqnarray}
    G_A^{Z,p} &=& -G_A^u + G_A^d +G_A^s \\
    G_A^{Z,n} &=& -G_A^d + G_A^u +G_A^s \> .
\end{eqnarray}
One can then isolate the strange axial form factor by
\begin{equation}
    G_A^s = \frac{(G_A^{Z,p} + G_A^{Z,n})}{2} \>.
\end{equation}
In principle, this would be a way to access the matrix element $G_A^s
(Q^2 =0) = \langle {\bar s} \gamma^\mu \gamma_5 s \rangle S_\mu$,
where $S$ is the nucleon spin four-vector. Again, in practice one must
consider radiative corrections and, for this axial form factor, the
contribution of anapole effects as discussed in \ref{sec:axial} below. The
theoretical uncertainties associated with these corrections render
measurement of the strange contribution $\Delta s$ problematic in parity
violating electron scattering.

For the vector form factors, the use of isospin symmetry yields the important relation
\begin{equation}
G_{E,M}^{Z,p} = (1 - 4 \sin^2 \theta_W) G_{E,M}^{\gamma,p}
		- G_{E,M}^{\gamma,n} - G_{E,M}^{s} \label{eq:EMZ}
\end{equation}
which illustrates how measurement of the electromagnetic form factors
for the proton and neutron, combined with measurement of the
corresponding $G^{Z,p}$, can provide access to the strange vector form
factors. In this case, as discussed in \ref{sec:rcorr}, the radiative
corrections are manageable and parity-violating electron scattering is
a very useful tool for studying strange vector form factors. In the
static limit, one has the particularly interesting quantity $\mu_s
\equiv G_M^s (Q^2 = 0)$ known as the strange magnetic moment. Since
the net strangeness in the nucleon is zero, one has $G_E^s (Q^2 = 0) =
0$. However, the mean squared strangeness radius $r^2_s \equiv -6
[{dG_E}/{dQ^2}](Q^2=0)$ is a static property that in general can be
non-zero.

Over the last 2 decades, there have been numerous theoretical papers
reporting predictions for strange vector form factors. Most of these
papers involve models of nucleon structure which, although
well-motivated, involve many uncertainties related to the expected
accuracy of the predictions. A good review of the model calculations
can be found in \cite{BeckHolstein}. Many of the models predict values
for $\mu_s$ and the results generally range from $-0.5 < \mu_s
<0.3$~n.m., with a strong preference for negative values. The
predicted values for $r_s$ also cover a substantial range $-0.25 < r_s
< 0.4$~fm.

In the last few years, calculations based on input from lattice QCD
methods have become available.  The challenge associated with lattice
calculations involves the evaluation of so-called ``disconnected''
insertions, where the vector current couples to a quark loop that does
not involve the valence quarks. (Since there are no valence strange
quarks, these are the amplitudes relevant to strange form factors). In
one approach \cite{Leinweber1}, the baryon octet matrix elements are
written with connected and disconnected insertions in separate terms.
Linear combinations of the baryon magnetic moments are used, along
with the assumption of charge symmetry, to obtain expressions
involving the baryon magnetic moments and ratios of amplitudes to be
evaluated using lattice calculations. It is claimed that the ratios
are reliably determined from lattice calculations, and this method
yields the prediction $\mu_s = -0.046 \pm 0.019$~n.m., with the
uncertainty estimated from the lattice statistical precision. A
similar approach \cite{Leinweber2} was then used to compute the
strange radius $r^2_s = 0.021 \pm 0.063$~fm$^2$, where the uncertainty
is predominantly due to the poor experimental information on baryon
charge radii.

More recently, a more direct approach \cite{Doi} using lattice methods
has been employed. Using a full QCD calculation using $N_f = 2+1$
clover fermion configurations the result $\mu_s = -0.017 \pm 0.025 \pm
0.07$~n.m. is obtained, where the first uncertainty is statistical and
the second is due to uncertainties in $Q^2$ and chiral extrapolations.
This result is in remarkable agreement with the more phenomenological
result \cite{Leinweber1}.

\subsection{Radiative corrections}
\label{sec:rcorr}
 In order to extract contributions of strange form factors $G^s$ from
 measurements of electroweak form factors, one must include the
 effects of ${\cal O}(\alpha)$ electroweak radiative
 corrections. These radiative effects typically arise from $\gamma-Z$
 ``box'' diagrams or loop effects.  It is common to express these
 ${\cal O}(\alpha)$ corrections as ratios $R_{V,A}$ (for vector $V$
 and axial vector $A$) which are fractions of the corresponding
 tree-level amplitudes. $R_V^p$, $R_V^n$, $R_V^{(0)}$ denote the
 ratios for vector proton, neutron, and SU(3)-singlet amplitudes,
 respectively. In principle, their values can be obtained using the
 Standard Model predictions for the effective electron-quark couplings
 $C_{1q}$ given in \cite{pdg}. However, these $C_{1q}$ do not include
 perturbative QCD contributions, nor coherent strong interaction
 effects in the radiative corrections associated with elastic
 scattering from a nucleon. A recent analysis of these effects has
 been given in Ref.~\cite{Erler:2003yk} and up-dated in
 Ref.~\cite{Erler:2004in} (which also includes an improved treatment
 of strong interaction contributions to the running of the weak mixing
 angle in the $\overline{MS}$ renormalization scheme from its value at
 the $Z$-pole) with the results
\begin{eqnarray}
     R_V^p   &=& -0.0520 \\
      R_V^n   &=& -0.0123  \\
      R_V^{(0)} &=& -0.0123 \> .
\end{eqnarray}
 The theoretical uncertainties in $R_V^{n}$ and $R_V^{(0)}$ are less
 than one percent, while the theoretical uncertainty in
 $(1-4\sin^2\theta_W)(1+R_V^p)$ is $\pm 0.0008$ \cite{Erler:2004in},
 or slightly more than one percent. (Since this error receives roughly
 equal contributions from the uncertainty in $\sin^2\hat\theta_W(M_Z)$
 as determined at the $Z$-pole and from the ${\cal O}(\alpha)$
 $Z\gamma$ box graph corrections, it is not appropriate to quote an
 uncertainty in $R_V^p$ alone.)  For the range of $Q^2$ associated
 with the experiments discussed in this review, the $R_V$ have a
 negligible impact on the $Q^2$-dependence of $A_{PV}^p$ and are taken
 to be constant.  We adopt the conventional $\overline{MS}$
 renormalization scheme so that $\sin^2\hat\theta_W$ is evaluated as
 $\sin^2\hat\theta_W(M_Z)= 0.231 16 \pm 0.00013$ \cite{pdg}.

\subsection{Axial Form Factor Corrections}
\label{sec:axial}

For the axial form factor, it is useful to employ the notation $G_A^e$
to differentiate the quantity relevant to parity-violating electron
scattering from other axial form factors (i.e., charged current
processes or neutrino scattering).  At lowest order this axial form
factor is the same as $G_A$ as measured in charged current
processes [$G_A (Q^2 = 0) = -1.2701 \pm 0.0025$] \cite{pdg}. 
However, the presence of strange quarks (i.e., the
contribution $\Delta s$) and radiative effects must be included, and
can be expressed
\begin{eqnarray}
  \label{eq_GAe}
\nonumber
  \displaystyle G_A^e(Q^2) &=& G_D(Q^2)\times
  [G_A (1+R_A^{T=1})+\frac{3F-D}{2}R_A^{T=0} \\
   &\ \ \ \ & +\Delta s (1+R_A^{(0)})]\,,
\end{eqnarray}
where
\begin{equation}
  G_D(Q^2) = \frac{1}{(1+Q^2/M_A^2)^2}\,,
\end{equation}
parameterizes the $Q^2$-dependence with a dipole form with the squared
axial mass $M^2_A = 1.00 \pm 0.04$~GeV$^2$ \cite{BBA03}. $F$ and $D$
are the octet baryon beta-decay parameters, which are determined from
neutron and hyperon beta decays under the assumption of SU(3) flavor
symmetry ($3F-D = 0.58 \pm 0.12$ \cite{DIS}).  $\Delta s = -0.07 \pm
0.06$ \cite{SMC} is the strange quark contribution to nucleon spin
obtained from inclusive polarized deep-inelastic lepton-nucleon
scattering.

The ratios $R_A^{T=1}$, $R_A^{T=0}$, and $R_A^{(0)}$ characterize the
effect of electroweak radiative corrections to the isovector,
isoscalar, and SU(3) singlet components of the axial form factor.
These quantities are traditionally divided into ``one-quark'' and
``many-quark'' contributions. The one-quark contributions correspond
to renormalization of the effective vector electron-axial vector quark
couplings, $C_{2q}$, and their values can be obtained from the
Standard Model predictions for these couplings given in
Ref.~\cite{pdg}.  The many-quark contributions include the so-called
``anapole'' effects as well as coherent strong interaction
contributions to the radiative corrections. In contrast to the vector
corrections, $R_V$, the {\em relative} importance of many-quark
effects in the $R_A$ can be quite pronounced.  The many-quark effects
can be addressed using chiral perturbation theory, and a comprehensive
analysis of the anapole contributions to $R_A^{T=1}$ and $R_A^{T=0}$
has been carried out to chiral order $p^3$ in Ref.~\cite{Zhu00}. The
total axial corrections, up-dated for the present value of the weak mixing
angle, are
\begin{eqnarray}
     R_A^{T=1}   &=& -0.258 \pm 0.34 \\
      R_A^{T=0}   &=& -0.239 \pm0.20  \\
      R_A^{(0)} &=& -0.55 \pm 0.55 \> .
\end{eqnarray}

\section{PARITY-VIOLATING ELECTRON SCATTERING}

\subsection{Theory}

The scattering of an electron from a hadronic target involves the
dominant electromagnetic amplitude due to photon exchange ${\cal
  M}_\gamma$ and the much smaller (at low momentum transfer $Q^2 \ll
M_Z^2$) neutral weak amplitude due to $Z$ exchange ${\cal M}_Z$. The
scattering cross section is related to the squared modulus of the sum of
these amplitudes $| {\cal M}_\gamma + {\cal M}_Z|^2$. Parity-violating
observables arise from the fact that the weak amplitude involves both
vector and axial vector currents, leading to pseudoscalar
quantities. The incident electron helicity, ${\hat s} \cdot {\hat k}$,
is a pseudoscaler quantity and so the helicity dependence of the cross
section violates parity symmetry and must involve the weak amplitude.
To lowest order, one expects the difference between positive helicity
and negative helicity cross sections to depend upon the product $d
\sigma_R - d \sigma_L \propto {\rm Re} [{\cal M}_\gamma {\cal
    M}^{VA}_Z]$, where ${\cal M}^{VA}_Z$ is the weak amplitude
associated with the product of vector and axial vector currents. The
helicity independent cross section is just due to the dominant photon
exchange amplitude $d \sigma_R + d \sigma_L \propto |{\cal
  M}_\gamma|^2$. Thus the parity-violating helicity-dependent
asymmetry has the structure
\begin{eqnarray}
  A_{LR} &\equiv& \frac{d \sigma_R - d \sigma_L} {d \sigma_R + d \sigma_L} \\
   &\propto& \frac{{\rm Re} [{\cal M}_\gamma {\cal M}^{VA}_Z]}{|{\cal M}_\gamma|^2}
\end{eqnarray}
The squared electromagnetic amplitude must be proportional to
$(e/Q^2)^4 = (4 \pi \alpha /Q^2 )^2$, whereas the product ${\cal
  M}_\gamma {\cal M}^{VA}_Z$ will be proportional to $(e/Q^2)^2
G_F/\sqrt{2} = 4 \pi \alpha G_F /\sqrt{2} Q^2$.  (At low $Q^2$ the
weak amplitude involves the Fermi coupling constant $G_F = 1.166 37
\times 10^{-5}$ ~GeV$^{-2}$.) Therefore, the parity-violating
asymmetry can be written
\begin{equation}
    A_{LR} = -{G_FQ^2\over 4 \sqrt{2}\pi\alpha}\times \frac {\cal N} {\cal D}
\end{equation}
in which the numerator ${\cal N}$ involves products of electromagnetic
and weak form factors and the denominator ${\cal D}$ involves squares
of electromagnetic form factors.

In lowest order, the expression for parity-violating electron nucleon scattering in
the laboratory frame of reference (where the initial nucleon is at rest) is given by
\begin{eqnarray}
    A_{LR} &\equiv& \frac{d \sigma_R - d \sigma_L}{d \sigma_R + d
    \sigma_L} \\
    &=& \left[\frac{-G_FQ^2}{4\sqrt{2}\pi\alpha}\right]
      \cdot (A_E+A_M + A_A) \label{eq:Nasym}
\end{eqnarray}
where the three terms are
\begin{eqnarray}
A_E &=& \frac{\epsilon G_E^\gamma G_E^Z }
             {\epsilon (G_E^\gamma)^2+ \tau (G_M^\gamma )^2}  \\
      A_M &=& \frac{ \tau G_E^\gamma G_M^Z}
             {\epsilon (G_E^\gamma)^2+ \tau (G_M^\gamma )^2}  \\
      A_A &=& \frac{-\epsilon^\prime ( 1-4 \sin^2 \theta_W) G_M^\gamma G_A^e}
            {\epsilon (G_E^\gamma)^2+ \tau (G_M^\gamma )^2}  \> ,
\end{eqnarray}
involving the kinematic variables
\begin{eqnarray}
\tau & = & {{Q^2} \over {4 M_N^2}} \nonumber \\
\epsilon & = & {1 \over 1 + 2(1 + \tau)\tan^2{\theta \over 2}} \nonumber \\
\epsilon^{\prime} & = & \sqrt{\tau (1+\tau) (1- \epsilon^2)}
\end{eqnarray}

which are functions of the momentum transfer $Q^2 = -q^2 >0$ and the
electron scattering angle $\theta$.  The different terms $A_E$, $A_M$
and $A_A$ depend upon the neutral weak form factors associated with
the electric, magnetic, and axial couplings to the nucleon and so
provide access to the strange form factors $G_E^s$, $G_M^s$, and
$G_A^s$, respectively. At small scattering angles, $\theta \rightarrow
0$ one finds that $\epsilon \rightarrow 1$, $\tau \rightarrow 0$, and
$\epsilon^\prime \rightarrow 0$, so $A_E$ becomes the dominant
term. At backward angles $\theta \rightarrow \pi$, $\epsilon
\rightarrow 0$ and the asymmetry is dominated by the magnetic and
axial asymmetries. Note also that the axial asymmetry is suppressed by
the factor $(1-4 \sin^2 \theta_W)$, but the actual numerical value is
quite sensitive to radiative and anapole effects. As a result, the
dependence of $A_{LR}$ on $G_A^s$ is quite small and so one uses
$G_A^s$ as determined from spin-dependent deep inelastic
scattering. Thus, in the end, one can treat $A_{LR}$ as a (linear)
function of the vector strange form factors $G_E^s$ and $G_M^s$.

By evaluating Eq.~\ref{eq:Nasym} in terms of the vector strange form factors
and including radiative corrections, one can express the asymmetry as
\begin{equation}
  \label{eq_linear_comb}
  A_{LR} = A_{nvs} + \eta_E G_E^s + \eta_M G_M^s\,,
\end{equation}
where $A_{nvs}$ is the ``non-vector-strange'' asymmetry (independent
of $G_E^s$ and $G_M^s$), and $\eta_E$ and $\eta_M$ are functions
of kinematic quantities and nucleon electromagnetic form factors.
For elastic scattering from the proton, one can measure the asymmetry at
a variety of scattering angles and fixed momentum transfer (by also varying the
incident beam energy) to obtain values for different linear combination of the
strange vector form factors. This procedure (analogous to the ``Rosenbluth''
separation for determining form factors from cross section measurements) facilitates
determination of both $G_E^s$ and $G_M^s$.

Another technique for experimental measurement of $G_E^s$ involves
elastic scattering from ${}^4$He. The ${}^4$He nucleus has spin $S=0$
and isospin $I=0$, so the magnetic and axial form factors vanish. The
charge form factor for electromagnetic scattering from ${}^4$He is
proportional to the isoscalar combination $G_E^{\gamma, p} +
G_E^{\gamma, n}$. Similarly, the neutral weak form factor is
proportional to $G_E^{Z, p} + G_E^{Z, n}$. One can use
Eq.~\ref{eq:EMZ} to obtain a relation between the electromagnetic form
factor for ${}^4$He, $F^\gamma (Q^2)$, and the neutral weak form
factor $F^Z (Q^2)$:
\begin{equation}
    F^Z (Q^2) = -F^\gamma (Q^2) \times \left(  4 \sin^2 \theta_W + \frac{2 G_E^s}{G_E^{\gamma, p} + G_E^{\gamma, n}}\right)
\end{equation}
which then yields the parity-violating asymmetry for elastic $e-{}^4$He scattering:
\begin{equation} \label{eq:A4He}
    A_{PV}^{He} = \displaystyle{\frac{G_FQ^2}{4\pi\sqrt{2}\alpha}}\\
\times\left(4\sin^2 \theta_W
+ \frac{2G_E^s}{G_E^p+G_E^n}\right) \> .
\end{equation}
One should keep in mind that radiative corrections, as discussed in
\ref{sec:rcorr} and \ref{sec:axial}, lead to minor modifications of
Eq.~\ref{eq:Nasym} and Eq.~\ref{eq:A4He} and must be included for any
quantitative analysis.

Finally, we mention that quasielastic scattering from deuterium has
been employed to study the axial form factor $G_A^e$. As mentioned
previously, this is not useful for constraining $G_A^s$, but does
provide a method to test the calculations of the radiative corrections
and anapole contributions to $G_A^e$. In the static approximation, one
can treat quasielastic scattering from deuterium as simply the sum of
scattering from a free proton and a free neutron. In this case the
parity violating asymmetry has the form
\begin{equation}
A_{d} = -{G_FQ^2\over 4 \sqrt{2}\pi\alpha}\times
   {{\cal N}_n + {\cal N}_p \over
    {\cal D}_n + {\cal D}_p}
\end{equation}
where ${\cal N}_p$ (${\cal D}_p$) and ${\cal N}_n$ (${\cal D}_n$) are
the numerators (denominators) in Eq.~\ref{eq:Nasym} for the proton and
neutron, respectively. The asymmetry again can be expressed as a sum
of three terms $A_E+A_M + A_A$ corresponding to numerator expressions
involving electric, magnetic, and axial weak form factors. At backward
angles, the contribution of $A_E$ is negligible. The strange magnetic
form factor contribution to $A_M$ is reduced, due to the combination
$(G_M^{\gamma, p} + G_M^{\gamma, n}) G_M^s$ and the small value of the
nucleon isoscalar magnetic form factor, relative to the axial term
$A_A$. Thus at backward angles the deuteron asymmetry, when combined
with the proton asymmetry, can provide useful information on the
isovector part of the weak axial form factor $G_A^e$. Again, it is
important to include radiative corrections in any quantitative
analysis. In addition, the nuclear effects associated with the binding
of the two nucleons in deuterium must also be considered.

\begin{figure}
\centerline{\psfig{figure=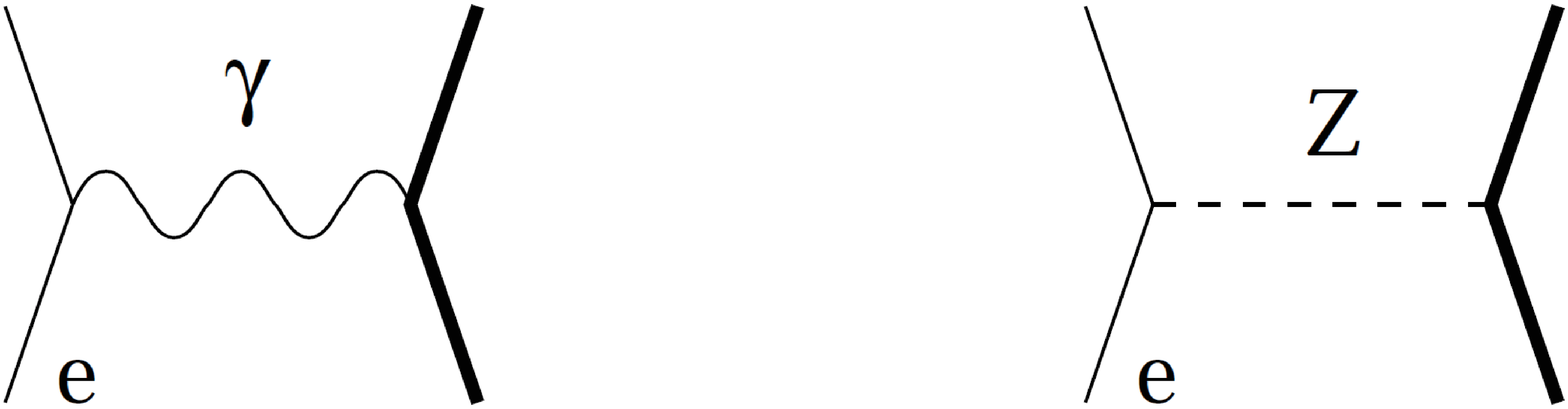,width=5.0in}}
\caption {The amplitudes relevant to parity-violating electron
scattering. The dominant parity-violating effects arise from
the interference of these two
amplitudes.}
\label{fig:amplitudes}
\end{figure}

\subsection{Experimental Technique}
The requirement to precisely measure party-violating asymmetries of
the scale of a few parts per million (ppm) imposes several significant
experimental challenges. High statistical precision demands long
running time and high luminosity, implying high beam current and thick
targets, as well as a highly-polarized beam. The beam polarizaton must
be well-determined, requiring high-quality beam
polarimetry. Backgrounds need to be well-separated from the elastic or
quasi-elastic scattering events of interest, and any residual
background contributions must be precisely corrected for. The beam
helicity needs to be rapidly flipped (typically at about 30 Hz) to
suppress the effect of slow changes in detector or beam properties. In
the ideal case, no other beam property should change when the helicity
is reversed.  Fluctuations in beam properties such as intensity,
trajectory, and energy that are correlated with the helicity ``flip''
must be minimized, and the sensitivity of the apparatus to such
helicity-correlated changes needs to be determined, so that
corrections can be made for the residual fluctuations. Additional slow
flips of the beam helicity can then be made to suppress many remaining
systematics. In this section we discuss several aspects of the various
approaches adopted to meet these challenges. \\

\subsubsection{Polarized Electron Source}

The need for a highly-polarized, high-current electron beam with
exquisite stability under reversal of the helicity dictates that a key
component of these experiments is the polarized electron source. The
adopted technology, pioneered at SLAC in the 1970's \cite{prescott},
is based on the emission of electrons from a GaAs photocathode when
exposed to circularly-polarized laser light. Beam polarizations of
near 40\% and currents of up to 100~$\mu$A were obtained using ``bulk''
GaAs crystals for the photocathode \cite{happex1_aniol1}.  Higher
polarization (at the expense of high beam current) was
subsequently produced using ``strained'' crystals, in which the active
layer of the photocathode was a thin ($\sim$100 nm) layer of GaAs
grown on GaAsP~\cite{happex1_aniol2}.
The mismatch between the two lattices produced a
strain in the GaAs, breaking the degeneracy in its energy levels,
theoretically allowing up to a 100\% polarization
~\cite{happex1_NAK91,happex1_MAR92}. More recently, the adoption of
multi-layer ``superlattice'' crystals~\cite{happex2_MAR04} has allowed both
high polarization (89\%) and high current (100~$\mu$A)
\cite{HAPPEXIII}.

\subsubsection{Beam Monitoring and Control}

High-precision parity-violation experiments impose stringent
requirements on the polarized electron beam in order to
minimize false asymmetries generated by helicity-correlated
variations in beam properties.
This demands careful attention
to the optical properties
of the incident laser light at the electron source.

The fast helicity reversal of the electron beam is accomplished by
reversing the handedness of the laser
light using a Pockels cell, which is a birefringent
crystal whose indices of refraction change with the application
of an electric field.
Linearly polarized light from the source laser, with
polarization at 45$^{\circ}$ to the transmission axes, acquires a phase
shift between the components on the slow and fast axes; by adjusting
the voltage
one can convert the light to either
right or left-handed circular polarization. Imperfections in the
Pockels cell typically lead to a small residual component of linear
polarization, which is different for the two nominal
circular polarization states.
 The linear components are transported
differently by the various optical elements, leading to
helicity-correlated intensity variations.  Gradients in the
birefringence in the cell also generate helicity-correlated changes in
the trajectory of the light, leading to electron beam position
variations ~\cite{Humensky:2002uv}, which can also lead to intensity
variations, as the quantum efficiency of the photocathode usually varies over
its surface.  The adoption of strained GaAs photocathodes makes these
effects especially acute: the strain introduces an anisotropy in the
quantum efficiency of the cathode, making it the dominant source of
analyzing power in the system.

Various passive and active techniques have been adopted to suppress
these helicity-correlated effects. Passive techniques include
careful
alignment of the laser beam through the Pockels cell \cite{Kent}, minimization of
optically active elements in the laser path, and attention to optimum
beam transport in the accelerator, to realize the natural ``adiabatic
damping'' of position fluctuations in the acceleration process. Active
techniques include feedback based on measurements of the
helicity-correlated changes in beam intensity and position in the
experimental hall.

\begin{figure}
\centerline{\psfig{figure=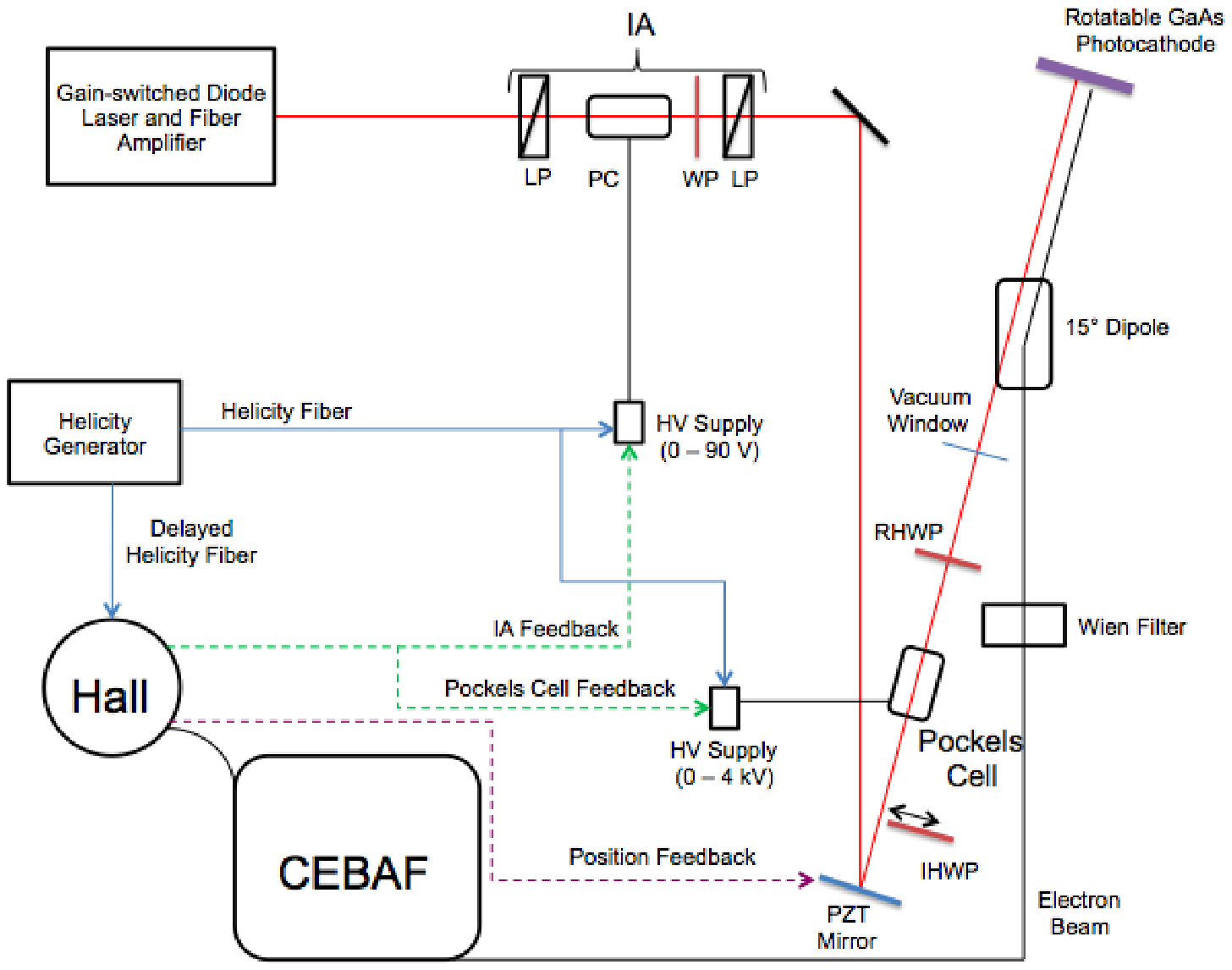,width=5.0in}}
\caption {Schematic of the polarized injector source at
  Jefferson Lab.  The laser light passes through an intensity
  attenuator system (IA) (see text) and  is directed by a
  PZT-mirror to the primary helicity-control Pockels cell.
  An insertable half-wave plate (IHWP) can be placed in
  the optical path (see text).  Passing through a
  rotatable half-wave plate (RWHP) the light is directed onto the
  photocathode. The emitted electron beam passes through a Wien filter,
  to adjust the spatial orientation of the electron
  spin, and is injected into the CEBAF accelerator.  The
  helicity generator signal is sent via fibre optics both to the
  high-voltage source for the Pockels cell and to the experimental
  data acquisition system.
  Helicity-correlated variations in the
 beam intensity and position are  monitored in the experimental Hall
  and can generate feedback signals to the source as outlined in the text.
 }
\label{fig:polsource}
\end{figure}

Figure~\ref{fig:polsource} depicts a typical polarized electron source
setup, in this case the Jefferson Lab source as used in recent parity
experiments \cite{Armstrong05, Aniol06a, Aniol06b, Acha07, Androic10}.
Two independent systems
have been used to supress the intensity
variations. The first
uses
adjustment of the voltage
signal to the Pockels cell - small differences in the voltages for the
two helicity states modify the residual linear polarization of the
transported light, leading to helicity-correlated intensity
variations, which then can be adjusted.  The second system, the IA
(Intensity Attenuator), consists of a second Pockels cell and a
waveplate (WP), sandwiched between two parallel linear polarizers (LP).
This
Pockels cell (PC) is operated at low helicity-dependent voltages (to
minimize beam-steering effects) and acts as an electro-optic
adjustable shutter. In either case, helicity-correlated electron beam
intensity variations are measured, tyically every few minutes,
and a feedback signal is sent to the electron source to
null the variations.

Both the SAMPLE experiment and forward-angle phase of the G0
experiment also used active feedback on helicity-correlated position
differences in the beam ~\cite{Averett:1999uy}. Here a
piezoelectric-controlled steering mirror (see
Fig.~\ref{fig:polsource}) is used to move the beam in a
helicity-controlled manner.

Another technique is to use an insertable half-wave plate
(IHWP) in the optical path, as shown in
Fig.~\ref{fig:polsource}. The IHWP is periodically inserted or
removed on a time scale of many hours. The IWHP rotates
the linear polarization state by 90$^{\circ}$, thereby inverting the
sense of the resulting circular polarization with respect to the
Pockels cell voltage. This slow ``flip'' should reverse the sign of
the measured asymmetry in the Hall with respect to the helicity
control signal, in the absence of any false asymmetry picked up in the
experimental electronics due to the helicity-control signal. Many
sources of helicity-correlated beam changes are insensitive to the
state of the IHWP, and thus this flip cancels these systematics.

To ensure that there are no false
asymmetries in the data due to the helicity-control signal being picked up
by the experimental electronics, the fast helicity reversal follows a
pseudorandom pattern, and the helicity state reported to the
electronics is delayed by several states in the pattern, and
only later reconstructed in software.

To illustrate the high suppresion of helicity-correlated beam
fluctuations afforded by these techniques, consider the HAPPEX-II measurement.
Here the values for these variations, averaged over the entire
several month-long run, were 0.4 ppm for intensity, $< 1.7$ nm
for position, $< 0.2$ nrad for angle and 0.2 ppb for energy.

The sensitivity of the apparatus to these
residual fluctuations can be determined either by multi-parameter
linear regression of the natural fluctuations, or by taking subsets of the
data with individual beam parameters modulated in a controlled manner
\cite{happex1_prc}.

\subsubsection{Beam Polarimetry}

Precision knowledge of the beam polarization is essential for normalizing
the measured asymmetries. Various polarimeters have been adopted in these
experiments, including those that measure continuously during the
experiment, such as transmission \cite{A4ComptonT,Beise05} and backscattering
\cite{HallACompton1, HallACompton2, HallACompton3, HallACompton4,A4ComptonB}
Compton polarimeters,
and those that are invasive, which can only make
periodic measurements of the polarization,
such as Mott \cite{Mott} and Moller \cite{HallAMoller, HallCMoller,A4Moller,Beise05}
polarimeters. The highest precision reached to date was in the HAPPEX-III
measurement, where a combination of Moller and backscattering Compton
devices yielded a precision of 0.75\% \cite{HAPPEXIII}.

\subsubsection{Counting vs. Integrating Methods for Asymmetry Measurements}
Two approaches have been adopted for the measurement of the parity-violating
asymmetry. The challenge is to accomodate the extremely high rates (of order 10
to 100 MHz) of scattered electrons that must be detected in order to achieve the desired
statistical precision. One approach is a counting method, in which
custom electronics are used to collect events in scalers (the G0 experiment)
or energy histograms (the PVA4 experiment) which are accumulated for a given
beam helicity state and then digitized at each helicity transition. In the
other approach, the integrating method (SAMPLE and HAPPEX experiments), the
analog signal from the detectors is integrated over a given helicity
window and the integral is digitized at each helicity change. Challenges
in the counting method include accounting for the effects of electronic
deadtime and event pileup which can distort the measured asymmetry, as well as
the design and cost of the custom electronics. In the integrating
method one has to ensure a high degree of linearity of the entire electronics chain
in order that helicity-correlated variations in the beam intensity not generate
false asymmetries. In neither approach, unlike the usual nuclear physics
experiment, can one digitize complete information
about individual scattering events, so only limited off-line analysis techniques
for dealing with background processes are available.

\subsubsection{Targets}

The program of parity-violating electron scattering experiments reviewed here generally
involve the use of targets of hydrogen, deuterium and helium. The very small asymmetries
to be measured imply that it is necessary to achieve high luminosity, so these targets
must be cryogenic to achieve sufficient areal density. The use of cryogenic fluids introduces
a new challenge: stability of the target density in the presence of the intense electron beam
which causes thermal heating of the fluid. Thermal fluctuations are a potential source of additional noise
that can degrade the statistical precision of the asymmetry measurement.

\subsubsection{Hydrogen and Deuterium Targets}

The operating point for a liquid hydrogen target is typically about 19
K, just a few degrees below the boiling point at a nominal pressure of
1-2~atm. Thus even with small temperature excursions the liquid
remains significantly below the boiling point to minimize thermal
fluctations due to the beam. Indeed localized boiling, leading to
bubble formation, is a major potential source of target thickness
variations that can contribute significantly to statistical noise in
the measured asymmetries. The fluid is generally pumped in a
recirculation loop in order to constantly provide fresh liquid within
the electron beam profile. The recirculation loop also contains a heat
exchanger where heat is transferred to helium gas as the primary
coolant.  An ohmic heater in the loop is used to control the
temperature and maintain the operating point by compensating for
variations in the beam power. Safety is a major consideration in the
design and operation of these targets, as a release of hydrogen gas to
the room atmosphere can lead to a dangerous flammable
mixture. Detailed descriptions of the SAMPLE \cite{SAMPLE_target}, G0
\cite{G0_target} and PVA4 \cite{A4_target} targets are available in
the literature.

\subsubsection{Helium Targets}

One experiment, HAPPEx-Helium, adopted a high pressure $^4$He gas
target to directly measureme the strange electric form factor $G_E^s$. 
The target was a 20-cm long cell
maintained at a temperature and pressure of 7 K and 13 atm, in which
the cryogenic $^4$He was pumped through the cell in a direction
transverse to the beam direction, so as to minimize local beam
heating.  Even without the possibility of bubble formation,
beam-induced density fluctuations in the gas were a challenge, but
were able to be limited to a few \% increase in the statistical noise
of the asymmetry measurement.

\section{EXPERIMENTAL RESULTS}

\subsection{SAMPLE}

\begin{figure}
\centerline{\psfig{figure=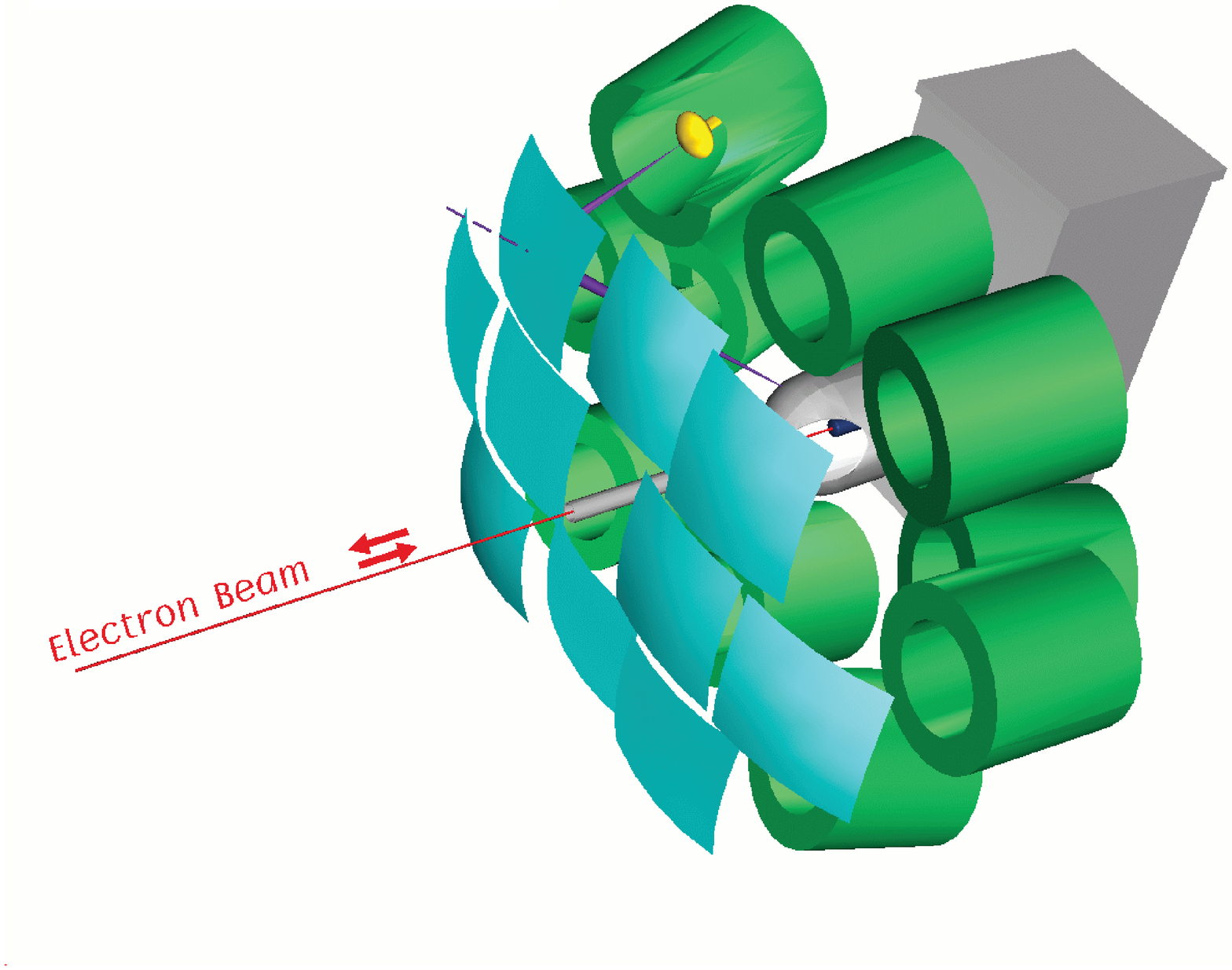,height=4.0in}}
\caption {Illustration showing the geometry of the SAMPLE target and
detector system. The electron beam is incident from the left. The electrons scattered at
backward angles emitted Cerenkov light in the air, which was reflected by 10 mirrors onto
ten 8-inch photomultiplier tubes. The photomultipliers were mounted inside cylindrical cast lead shields
to reduce background.}
\label{fig:sampleSchematic}
\end{figure}

The SAMPLE experiment, performed at the MIT-Bates Linear Accelerator
Center, was primarily focused on the goal of determining the strange
magnetic form factor $G_M^s$ at low momentum transfer $Q^2 =
0.1$~GeV${}^2$.  A rather complete description of the experiment and
results can be found in \cite{Beise05}, and we provide a short summary
here.  A 200~MeV polarized electron beam, typically 40~$\mu$A, was
incident on a 40~cm long liquid hydrogen target.  A schematic of the
apparatus is shown in Fig.~\ref{fig:sampleSchematic}. The scattered
electrons were detected in a large solid angle ($\sim 1.5$ sr) air
Cerenkov detector, with 10 ellipsoidal mirrors at backward angles
$130^\circ < \theta < 170^\circ$, resulting in an average $Q^2 \simeq
0.1$ (GeV/$c)^2$.  At these kinematics the axial term is expected to
contribute about 20\% of the asymmetry.

The hydrogen data set was acquired in 1998, and the experiment was run
with a deuterium target in 1999 to acquire data on the axial form
factor.  The initial deuterium result indicated a substantial
discrepancy with calculations of the axial form factor, and so further
deuterium experiments at lower momentum transfer were performed in
2000-2001. These experiments, along with a reanalysis of the 1999
deuteron dataset (with improved accounting of pionic backgrounds) now
provide significant confirmation of the theoretical treatment of the
axial form factor.

The polarized electron source utilized bulk GaAs, with polarization
typically about 35\%.  The Bates beam was pulsed at 600~Hz, and the
beam helicity was changed for each pulse. A preselected random pattern
of ten helicity states was generated, and the complement of this set
(with reverse helicity) followed for the next ten pulses.  The
longitudinal spin emerging from the polarized source was pre-rotated
by a Wien spin rotator before injection into the accelerator so the
the $36.5^\circ$ magnetic bend into the experimental hall would then
provide longitudinal polarization at the target location.  The beam
polarization was periodically measured using a M{\o}ller polarimeter.

Each detector signal was integrated over the
$\sim 25 \mu$s of the beam pulse and digitized along with the beam charge
for that pulse. The ratio of integrated
detector signal to integrated
beam charge (normalized yield) was then
corrected for background, beam polarization, and other systematic beam effects
to give the experimental result for the parity-violating asymmetry.

For the SAMPLE kinematics,
the parity-violating asymmetry for elastic
scattering on the proton for the incident electron energy of 200~MeV
can be written as
\begin{equation}
A_p
= -5.56 + 1.54G^e_A(T=1) + 3.37 G^s_M \> {\rm ppm}.
\end{equation}
(The isoscalar component of $G_A^e$ is computed to be very small
\cite{Zhu00} and we have
absorbed it into the leading constant term.)
The SAMPLE measurement of this asymmetry \cite{Spayde00}  is
\begin{equation}
A_p = -5.61 \pm 0.67_{stat} \pm 0.88_{sys} \> {\rm ppm}.
\end{equation}
Using the value of $G_A^e(T=1) = -0.83 \pm 0.26$ from \cite{Zhu00} results
in the strange magnetic form factor
\begin{equation}
    G^s_M(Q^2 = 0.1) = 0.37 \pm 0.20 \pm 0.26 \pm 0.07 \label{eq:Hresult}
\end{equation}
where the first uncertainty is statistical, followed by the estimated
experimental systematic uncertainty and the last uncertainty is due to
the axial corrections and electromagnetic form factors.

The asymmetry for quasielastic electron scattering deuterium
for the SAMPLE kinematics and detector acceptance
at 200~MeV is written
\begin{equation}
A_d
= -7.06 + 1.66 G^e_A(T=1) + 0.72 G^s_M \> {\rm ppm}
\end{equation}
where the corrections for deuteron structure and other nuclear
effects, including hadronic parity violation, have been included as
discussed in \cite{rocco04}.
One should note that the deuteron asymmetry is more sensitive to the
contribution from the isovector axial form factor $G^e_A(T=1)$
than the proton asymmetry \cite{Beise91}.
The updated SAMPLE result for the deuterium asymmetry in quasielastic
kinematics is \cite{Ito04}
\begin{equation}
A_d = -6.79 \pm 0.64_{stat} \pm 0.55_{sys} \> {\rm ppm} \> .
\end{equation}

A combined fit of the H and D data, assuming $G_E^s =0$,
allows a separation of $G_M^s$ and $G_A^{T=1}$ and
yields \begin{eqnarray}
  G_M^s &=& 0.23 \pm 0.36_{stat} \pm 0.40_{sys} \\
  G_A^e (T=1) &=& -0.53 \pm 0.57_{stat} \pm 0.50_{sys} \> ,
\end{eqnarray}
which agrees well with the H result (Eq.~\ref{eq:Hresult})
and also the theoretical prediction \cite{Zhu00} for $G_A^e (T=1)$.

\subsection{HAPPEX}

The HAPPEx series of experiments were run in Hall A at Jefferson Lab,
and made use of the High
Resolution Spectrometers \cite{A-NIM} to focus elastically scattered
particles onto a total absorption shower counter in each focal plane
whose signals were integrated over the 33 ms helicity period. The HRS
suppressed background from inelastic scattering and low-energy secondaries.

The first generation HAPPEX experiment ran in 1998 and 1999
\cite{happex1_aniol1,happex1_aniol2,happex1_prc} at a kinematics
$\langle \theta_{\rm lab} \rangle = 12.{3^\circ} $ and $\langle Q^2
\rangle = 0.477$ (GeV/c)$^2$ corresponding to the smallest angle and
largest energy possible with the Hall A HRS spectrometers, which
maximized the figure of merit for a first measurement.

In the 1998 run the experiment used a $I = 100 \mu$~A beam
with $P = 38$\% polarization produced from a
bulk GaAs crystal, while in the 1999 run
HAPPEX-I became the first experiment to use a
strained GaAs photocathode to measure a parity-violating
asymmetry in fixed-target electron scattering.
This improved the figure of merit $P^2 I$ with
$P$=70\% and $I$=35 $\mu$A.
\cite{happex1_prc}

During HAPPEX-I the Hall A Compton polarimeter \cite{HallACompton1}
was commissioned and provided, for the first time, a continuous monitoring of
the electron beam polarization with a total relative error from
run-to-run of less than 2\%.  The Compton results were in
good agreement with the M{\o}ller polarimeter in Hall A and a Mott polarimeter
located in the 50 MeV region of the accelerator.

The HAPPEX-I physics asymmetry was found to be
\begin{eqnarray}\label{happex1_asy}
A = -15.05 \pm 0.98_{stat} \pm 0.56_{sys} \; {\rm ppm}.
\end{eqnarray}
The precision of the result was sufficient to rule out several then-current
theoretical
estimates of strangeness effects at moderately high $Q^2$ where
it was thought the effects might have been large \cite{happex1_prc}.

Using this result, along with the calculated $G^{Z p}_A$ and the known
values of the proton and neutron form factors, the experiment determined the
linear combination of strange
form factors
\begin{equation}
\label{eq:happex1_strff}
G_E^s + 0.392 G_M^s = 0.014 \pm 0.020 \pm 0.010
\end{equation}
where the first error is the total experimental error
(statistical and systematic errors added
in quadrature) and the second error
is the error due to the
``ordinary'' electromagnetic form factors. One feature of the HAPPEX
experiments is that they have negligible sensitivity to the axial form factors,
whose effect is kinematically suppressed due to the very forward scattering
angle.

The second generation HAPPEX experiments, HAPPEX-II and
HAPPEX-Helium took data in 2004 and 2005 at a lower $Q^2 \sim 0.1$ GeV$^2$,
by utilizing superconducting
septa magnets to allow the HRS to detect elastically scattered
electrons at a scattering angle of $6^{\circ}$.
The measurement on the $^4$He target yielded an asymmetry of
\begin{equation}
A_{He} = +6.40 \pm 0.23_{stat} \pm 0.12_{sys} \> {\rm ppm} \> ,
\end{equation}
which tightly constrained the strange electric form factor to
$G^s_E = 0.002 \pm 0.014 \pm 0.007$. The hydrogen data asymmetry
\begin{equation}
A_{p} = -1.58 \pm 0.12_{stat} \pm 0.04_{sys} \> {\rm ppm} \>
\end{equation}
determined the form factor combination
\begin{equation}
G_E^s + 0.09 G_M^s = 0.007 \pm 0.011_{stat} \pm 0.006_{sys} ,
\end{equation}
again, consistent with zero.

Subsequently, the HAPPEX-III experiment returned to higher $Q^2$ (0.62 GeV$^2$) with a data-taking
run in 2009. This was motivated by indications of significant strange form-factor
contributions at high ($>$ 0.4 GeV$^2$) Q$^2$ in results from the forward-angle
phase of the $G^0$ experiment (see \ref{sec:G0}). HAPPEX-III capitalized on advacemements in
polarimetery, control of helicity-correlated beam fluctuations, and improved figure of merit
compared to HAPPEX-I, leading to a precision asymmetry measurement
\begin{equation}
  A_p= -23.80 \pm 0.78_{stat} \pm  0.36_{sys} \> {\rm ppm} \>,
\end{equation}
yielding the form factor combination
\begin{equation}
G_E^s + 0.517 G_M^s = 0.003 \pm 0.004_{stat} \pm 0.009_{sys} ,
\end{equation}
where the third error is due to electromagnetic form factors and radiative corrections. Again,
the result is consistent with zero net strangeness contribution.

\subsection{$G^0$}
\label{sec:G0}

The G0 experiment was performed in Hall C at Jefferson Lab. In this
experiment the parity-violating asymmetry in elastic electron
scattering from hydrogen and quasi-elastic electron scattering from
deuterium was measured in the $Q^2$ range 0.1-1 (GeV/c)$^2$ in both
forward and backward angle modes. By measuring three independent
asymmetries, one at forward angles on liquid hydrogen and two at
backward angles, one on liquid hydrogen and one on liquid deuterium, a
complete separation of the strange vector form factors of the nucleon
($G^s_M$, $G^s_E$) and the isovector axial form factor ($G^e_A(T =
1)$) was possible. In the forward angle mode the particle detected was
the recoil proton, while in the backward angle mode the particle
detected was the scattered electron. The experiment ran in forward
angle mode during the period 2002-5 and in backward angle mode during
2006-7.

\begin{figure}
\centerline{\psfig{figure=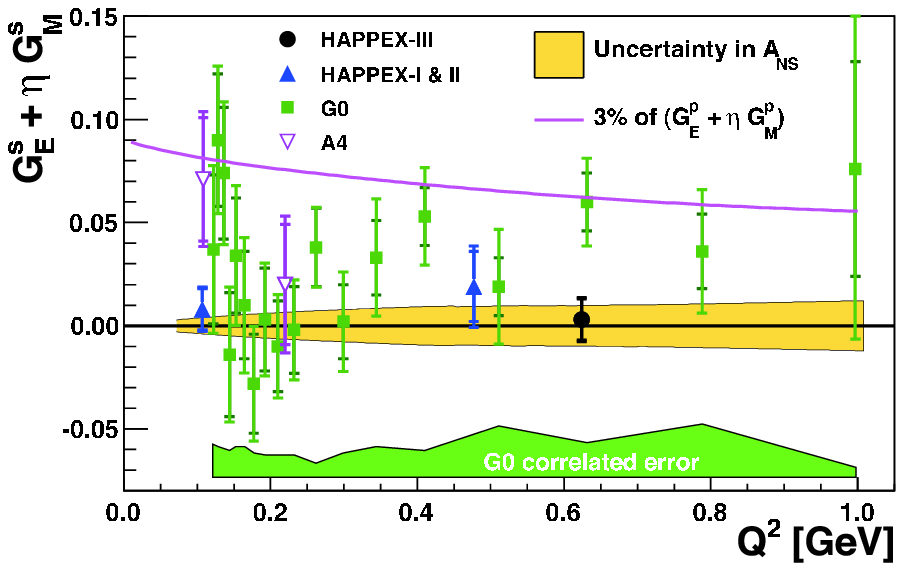,width=6.0in}}
\caption{Strange quark vector form factor results from all forward-angle
scattering measurements on the proton as a function of $Q^2$.
For the $G^0$ results \protect
\cite{Armstrong05}, the inner error
bars are statistical and the outer are point-to-point systematics, with the correlated
systematic error shown as a green band. The yellow band shows the uncertainty
in the predicted asymmetries in the absence of strangeness effects, due to knowledge
of the electromagnetic and axial form factors. For reference, the solid curve shows
a 3\% contribution to the comparable linear combination of proton form factors.}
\label{fig:HAPPEXIII}
\end{figure}

The G0 experiment employed a large-acceptance superconducting toroidal
spectrometer with eight coils and eight sets of particle detectors
providing excellent azimuthal symmetry about the beam axis. In forward
angle mode, the recoiling protons were detected using 16 pairs of
plastic scintillation detectors in each octant. Each detector pair
(one behind the other) selected coincident events to reduce accidental
backgrounds. The incident beam of 3~GeV electrons was delivered in
short ($\sim 100~$ps) pulses at 31~MHz to allow a 32~ns time of flight
window for detection of the recoil protons. Custom time-encoding
electronics enabled measuring the protons as a function of time of
flight in pulse counting mode.  At the larger $Q^2 \ge
0.3$~(GeV/c)$^2$, large positive asymmetries due to hyperon decay
backgrounds necessitated careful treatment and correction to extract
the much smaller negative asymmetries due to parity violation in
elastic scattering. The forward angle G0 measurements
\cite{Armstrong05} enable a simultaneous determination of the quantity
$G_E^s + \eta G_M^s$ over the $Q^2$ range 0.1-1~(GeV/c)$^2$, where
$\eta \equiv \tau G_M^\gamma / \epsilon G_E^\gamma$.  The results are
shown in Fig.~\ref{fig:HAPPEXIII}.

\begin{figure}
\centerline{\psfig{figure=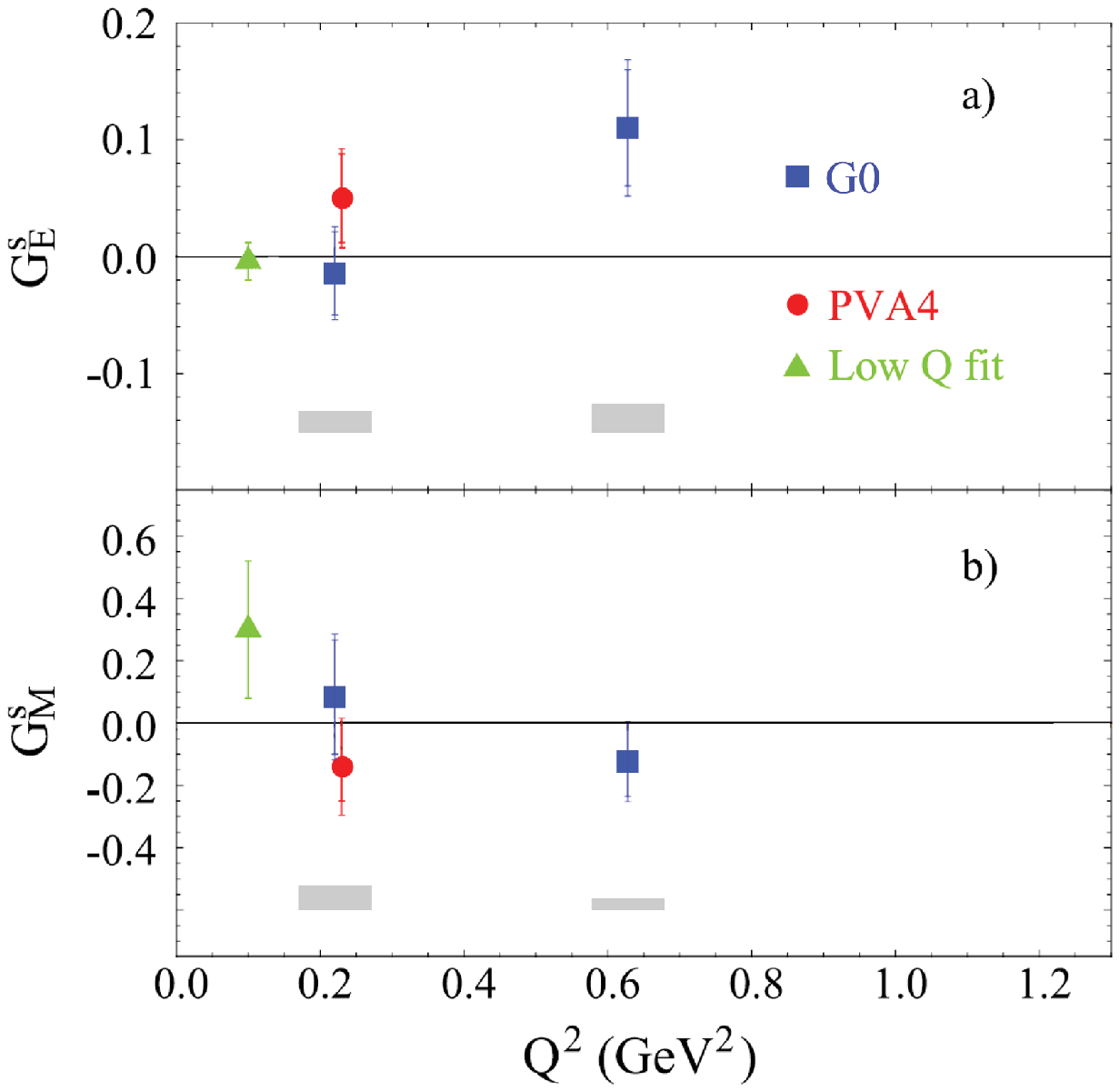,width=5.0in}}
\caption {Results for the strange vector form factors extracted
from combining the forward and backward-angle G0 experiments
on hydrogen and deuterium \cite{Androic10}; also
shown are the results from PVA4 at Q$^2 = 0.23$ GeV$^2$ \cite{Maas04,Baunack09},and a global fit \cite{Liu} to world data at $Q^2 = 0.1$ GeV$^2$. The
grey bands are correlated systematic errors for the G0 data. }
\label{fig:G0back}
\end{figure}

The G0 backward angle results were obtained using incident beams at
359~MeV and 684~MeV. The orientation of the toroidal spectrometer was
reversed to facilitate measurement of scattered electrons near
$110^\circ$ with respect to the incident beam direction. The
scintillation detectors (called focal plane detectors, FPD) were
supplemented with additional scintillators near the exit of the magnet
(cryostat exit detectors, CED) and aerogel threshold Cerenkov counters
(pion threshold 570~MeV) to reject pions. The backward angle results
(hydrogen and deuterium) were combined with the forward angle
measurements to yield values of $G_E^s$, $G_M^s$, and $G_A^e (T=1)$
\cite{Androic10} at two values of $Q^2$ (0.221 and 0.628 GeV$^2$) as
shown in Fig.~\ref{fig:G0back}.

\subsection{PVA4}

The PVA4 collaboration at the MAMI microton adopted a counting-mode
approach, with a highly segmented calorimeter along with custom fast
electronics.  No magnetic spectrometer is used; the signal is
separated from backgrounds using the energy deposition in the
calorimeter. The calorimeter is an azimuthally symmetric array of 1022
PbF2 crystal, in seven rings covering scattered electron angles from
either $30^{\circ}$ to $40^{\circ}$ (forward configuration) or
$140^{\circ}$ to $150^{\circ}$ (backward configuration), acting as a
total-absorption Cerenkov detector \cite{Baunack11}. The spectrum of
energy deposited above a hardware threshold in clusters of 9 crystals is histogrammed
using pipelined fast digitizer and the energy histograms are stored
for each helicity state.  The beam current is typically 20 $\mu$A,
with a polarization of 80\%, and the helicity state is selected every
20 ms.

The first PVA4 measurement was in a forward-angle configuration,
with a liquid hydrogen target; the measured asymmetry is sensitive to
a linear combination of $G^s_E$ and $G^s_M$.
The beam energy was 855 MeV yielding a $Q^2 =
0.230$ $($GeV/$c)^2$.  For this initial measurement, only half
of the 1022 detector channels were
instrumented.  The measured asymmetry was \cite{Maas04}.
\begin{equation}
A_p (Q^2 = 0.230) = -5.44 \pm 0.54_{stat} \pm 0.26_{sys}  \; \rm{ppm} \> .
\end{equation}
This asymmetry implies a value for the linear combination of the strange form factors
of \begin{equation}
(G^s_E + 0.225 G^s_M)(Q^2 = 0.230) = 0.039 \pm 0.034 .
\end{equation}

\begin{figure}
\centerline{\psfig{figure=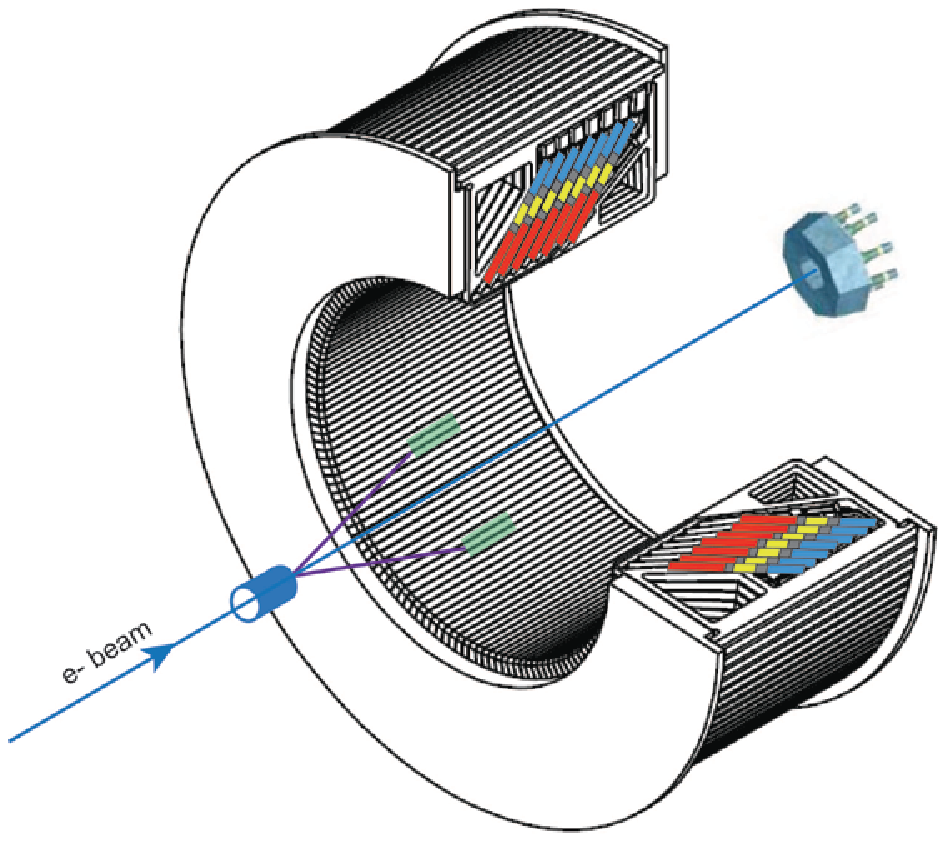,width=7.0in}}
\caption{Layout of the PVA4 detector. The electron beam is incident from the left;
scattered electrons are detected in the projective-geometry PbF2 calorimeter,
consisting of 7 rings of crystals in 146 rows.
Beam intensity fluctuations are monitored using a water Cerenkov luminosity monitor system detecting small-angle
scattering, shown on the right.}
\label{fig:PVA4}
\end{figure}

The second PVA4 forward-angle measurement made use of the fully
instrumented detector.
Data were taken on
liquid hydrogen with
a beam energy of 570 MeV, at $Q^2 = 0.108$ $($GeV/$c)^2$, and yielded
an asymmetry of
\begin{equation}
A_p (Q^2 = 0.108) = -1.36 \pm 0.29_{stat} \pm 0.13_{sys}  \; \rm{ppm},
\end{equation}
which implies the value of
\begin{equation}
(G^s_E + 0.106 G^s_M)(Q^2 = 0.108) = 0.071 \pm 0.036
\end{equation}
for the linear combination \cite{Maas05}.  This later result
represents a nearly 2$\sigma$ deviation from zero.

After these two forward-angle measurements, the PVA4 apparatus was
turned around and modified for a series of backward-angle
measurements.  Added to the detector was a double-ring of 72
scintillator counters, each of which cover 14 of the PbF$_2$
detectors.  These are used for electron tagging in order to suppress
the copious background of photons arising from $\pi^0$ decay.

The first back-angle measurement on hydrogen was at 315 MeV beam energy and
 $Q^2$ of 0.230 (GeV/$c$)$^2$, to match the $Q^2$ of one of the
forward-angle points.
The result was \cite{Baunack09}
\begin{equation}
A_p (Q^2 = 0.230) = -17.23 \pm 0.82_{stat} \pm 0.89_{sys}  \; \rm{ppm} \> .
\end{equation}
The measured
asymmetry implies a value for the linear combination of the strange form factors
of \begin{equation}
(G^s_M + 0.26 G^s_E)(Q^2 = 0.230) = -0.12 \pm 0.11_{stat} \pm 0.11_{sys} .
\end{equation}

Data were also taken at the same kinematics with a liquid deuterium
target; results are expected soon \cite{A4Progress}. The MAMI accelerator has undergone an
energy upgrade to allow beam energies up to 1.5 GeV, opening up a
wider range of $Q^2$ for the PVA4 experiment. The collaboration has
since moved the detector back into its forward scattering
configuration \cite{A4Progress}, and taken data at $Q^2$ = 0.63 (GeV/$c$)$^2$, which
matches the kinematics of both the HAPPEx-III experiment and the
higher $Q^2$ point at backward angle from G0.

\section{CONCLUSIONS}

During the last decade there has been a substantial international
effort to perform measurements of parity-violating asymmetries in
elastic electron scattering from nucleons. The primary aim of this
program has been to constrain the strange quark-antiquark
contributions to the nucleon electroweak form factors, $G_E^s$ and
$G_M^s$, as a function of momentum transfer $Q^2$.  The experiments
have made great progress in advancing the techniques required to
perform reliable and precise measurements that enable extraction of
the form factors. Substantial theoretical effort has provided
confidence in the radiative corrections and the degree of uncertainty
associated with contributions to the axial form factor $G_A^e$. As a
result, we now have a rather clear picture that has emerged from this
body of work.

In general, a convincing signal for a significant strange
quark-antiquark effect in the vector form factors has not been
obtained from these measurements. The various experimental results at
different kinematics all seem to support this general conclusion. In
fact, the results are remarkably consistent with this conclusion
despite the difficulty of these very challenging experiments.

\begin{figure}
\centerline{\psfig{figure=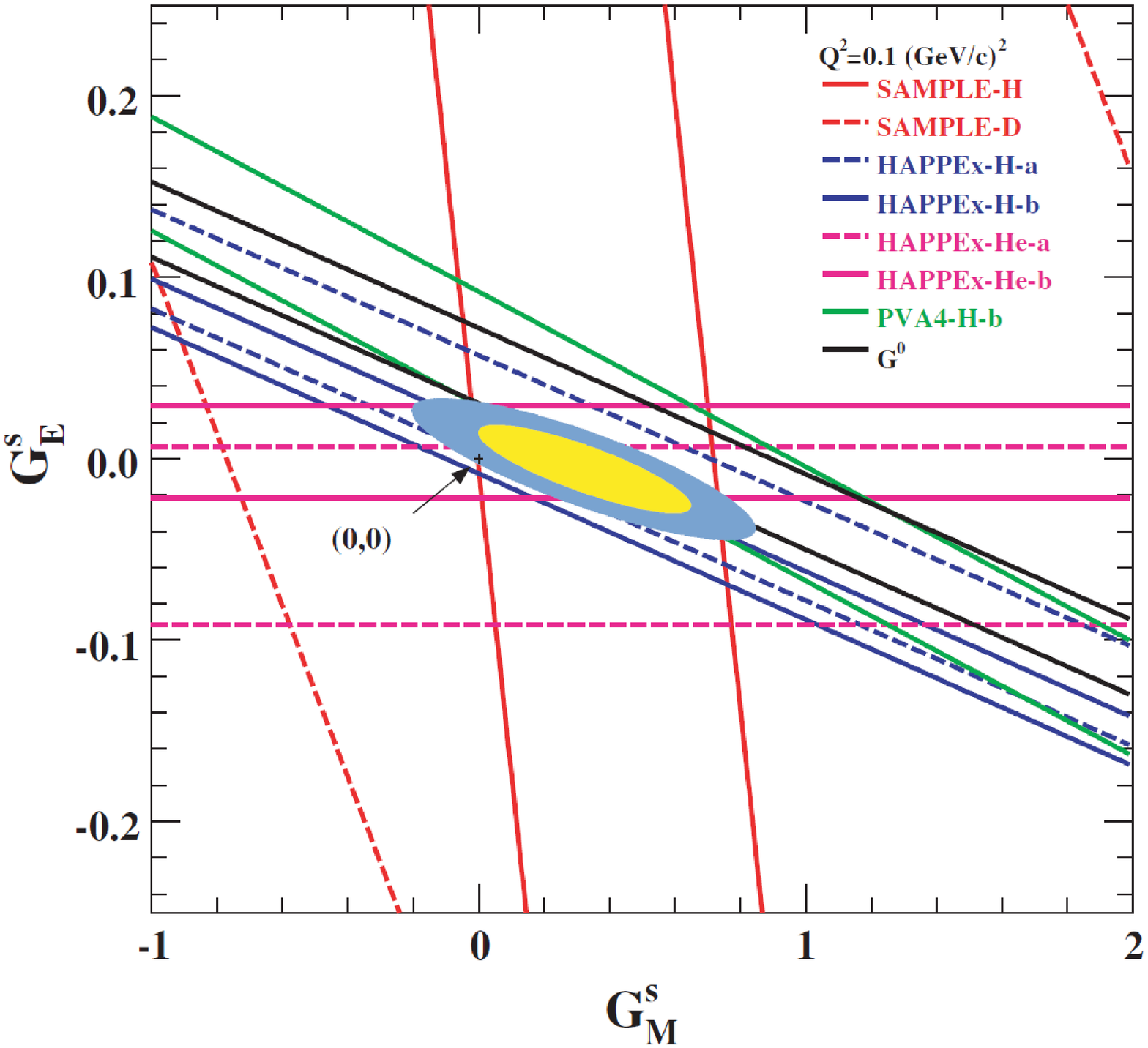,width=5.0in}}
\caption {The world data constraints on ($G^s_E,G^s_M$)
at $Q^2 = 0.1$(GeV/$c)^2$. The form factors of Kelly are used. Different
bands in the plot represent SAMPLE-H \cite{Spayde00} (solid red), SAMPLE-D
\cite{Ito04} (dashed red), HAPPEx-H-a \cite{Aniol06a} (dashed blue), HAPPEx-H-b
\cite{Acha07} (solid blue), HAPPEx-He-a \cite{Aniol06b} (dashed pink), HAPPEx-He-b
\cite{Acha07} (solid pink), PVA4-H-b \cite{Maas05} (solid green), and the lowest three
$Q^2$ bins in $G^0$ forward angle \cite{Armstrong05} (solid black). The yellow and
gray blue (dark) ellipses represent 68.27\% ($\Delta \chi^2 = 2.3$) and 95\%
($\Delta \chi^2 = 5.99$) confidence contours around the point of maximum
likelihood at ($G^s_E= 0.006$, $G^s_M
= 0.33$). The black cross represents
$G^s_E = G^s_M
= 0$.}
\label{fig:Liu}
\end{figure}

At the lowest momentum transfers ($Q^2 \sim 0.1$~(GeV/c)${}^2$, there
have been global fits performed to the body of experimental data in
this kinematic region \cite{Liu, Young}. These fits illustrate the
consistency of the data and reinforce the conclusion that the strange
vector form factors are quite small (compared to many model
predictions) at this low $Q^2$. Fig.~\ref{fig:Liu} shows the result of
the most recent of these global fits \cite{Liu}. As a result, one can
now conclude that, with 95\% confidence, strange quarks contribute
less than 5\% of the mean-square charge radius and less than 6\% of
the magnetic moment of the proton.

It is fair to say that these results are rather surprising in light of
the guidance from many models of nucleon structure that predicted
substantial strange quark effects at low $Q^2$. The result also seems
to be at variance with the notion that baryon-meson fluctations are a
significant aspect of nucleon structure as one would infer from the
${\bar d}-{\bar u}$ asymmetry observed in Drell-Yan production and the
success of many meson cloud models of the nucleon. More recently,
there have been results based on Lattice QCD calculations
\cite{Leinweber1, Leinweber2, Doi} that seem to support the very small
values of the strange form factors indicated by the experiments. It is
certainly gratifying that these calculations are consistent with
experiment. Nevertheless, they do not provide much insight as to {\it
  why} the strange quark contributions are suppressed in these
quantities. In the end, one has to admit that we have certainly
learned something quite significant about nucleon structure from this
program but that a deeper understanding of this phenomenon is still
lacking.

On the practical side, the conclusion that strange form factors are
constrained to be small combined with the development of the
experimental techniques for parity-violating electron scattering
experiments has motivated new experiments to perform precision tests
of the standard electroweak model in parity-violating electron
scattering. In particular, the experimental program reviewed here, and
the constraint on the strange electric form factor of the proton
$G_E^s$, provides a quantitative basis for assessing the utility of
low $Q^2$ measurements of parity-violating electron scattering to
provide precise new information on the running of the weak mixing
angle $\theta_W$ \cite{carlini}.  For example, the Qweak experiment at
Jefferson Lab \cite{Qweak,vanOers} is presently underway and expects to
provide a measurement of $\sin^2 \theta_W$ to about 0.24\%
precision. Such a measurement would herald a new era of precision
tests of the standard model that could reveal effects associated with
new physics at the TeV scale.

\section{ACKNOWLEDGEMENTS}
This work was supported in part by DOE contract DE-AC05-06OR23177, under 
which Jefferson Science Associates, LLC, operates the Thomas Jefferson National
Accelerator Facility, and by NSF grants PHY-0758068 and PHY-1068667.
We also thank our colleagues in the SAMPLE, HAPPEX, G0, and PVA4 collaborations 
for many fruitful discussions and interactions. 

\section{NUMBERED LITERATURE CITED}

\end{singlespace}

\end{document}